# Contextualization of topics - browsing through terms, authors, journals and cluster allocations

Rob Koopman[1] , Shenghui Wang[2] and Andrea Scharnhorst[3]

[1] *rob.koopman@oclc.org*
OCLC Research, Schipholweg 99, Leiden (The Netherlands)
[2] *shenghui.wang@oclc.org*
OCLC Research, Schipholweg 99, Leiden (The Netherlands)
[3] *andrea.scharnhorst@dans.knaw.nl*
DANS-KNAW, Anna van Saksenlaan 51, The Hague (The Netherlands)

## Abstract

This paper builds on an innovative Information Retrieval tool, *Ariadne*. The tool has been developed as an interactive network visualization and browsing tool for large-scale bibliographic databases. It basically allows to gain insights into a topic by contextualizing a search query (Koopman et al., 2015). In this paper, we apply the Ariadne tool to a far smaller dataset of 111,616 documents in astronomy and astrophysics. Labeled as the *Berlin dataset*, this data have been used by several research teams to apply and later compare different clustering algorithms. The quest for this team effort is how to delineate topics. This paper contributes to this challenge in two different ways. First, we produce one of the different cluster solution and second, we use *Ariadne* (the method behind it, and the interface - called *LittleAriadne*) to display cluster solutions of the different group members. By providing a tool that allows the visual inspection of the similarity of article clusters produced by different algorithms, we present a complementary approach to other possible means of comparison. More particular, we discuss how we can - with *LittleAriadne* - browse through the network of topical terms, authors, journals and cluster solutions in the *Berlin dataset* and compare cluster solutions as well as see their context.

## Introduction

What are essence and boundary of a scientific field? How can a topic be defined? Those are questions that are core to bibliometrics. Rigour and stability in defining boundaries of a field are important for research evaluation and funding distribution. However, if you as a researcher would seek for information about a certain topic of which you are not an expert yet, your information needs are quite different. Among the many possible hits for a search query you might want to know which are core works (articles, books) and which are rather peripheral. You might want to use different rankings (Mutschke & Mayr, 2014) or get some context. On the whole you would have less need to define a topic and a field in a bijective, univocal way. The same holds if you want to compare different clustering algorithms. Here again, you are in need to illustrate similarities and differences between different allocations of documents to clusters. Ways to contextualize them and browse through these contexts would be desirable. This is our starting point.

Decades of bibliometrics research have produced many different algorithms to cluster bibliographic records. They often focus on one entity of the bibliographic record. For example, articles and terms those articles contain (in title, abstract and/or full text) form a bipartite network from which we can either build a network of related terms (co-word analysis) or a network of related articles (based on shared words). The first method, sometimes also called lexical, has been often applied in scientometrics to produce so-called topical or semantic maps. The same exercise can be applied to authors and articles, articles and journals, in effect each element of the bibliographic record for an article. (Havemann, Scharnhorst 2012) If we extend the bibliographic record with the list of references, we enter the area of citation analysis. Here two methods are widely used: direct citations (known as delivering often sparse matrices) and co-citation maps (known as a good method to identify research fronts). Hybrid methods combine citation and lexical analysis (e.g., Zitt, Bassecoulard 2006; Janssens et al. 2009). The majority of studies applies one technique. But, sometimes analysis and visualization of multi-partite networks can be found (cf. Van Heur, Leydesdorff, & Wyatt 2013).

Each of the possible different network representations of articles stands for another aspect of connectivity between published scientific work. Co-authorship networks shed light on the social dimension - the invisible colleges - of knowledge production (Mali et al. 2012; Glänzel, Schubert 2004). Citation relations are interpreted as traces of flows of knowledge (Price, 1965; Radicchi et al. 2012). Depending on which element of the bibliographic record is used, we obtain different perspectives on how a field or a topic is to be conceived - as conceptional, cognitive unit; as a community of practice; or as institutionalized in journals. We can call this a *measurement effect*. Another source of variety next to differences resulting from what to analyze is how to analyze it. Finding clusters is part of network analysis. But, clusters can be defined in different ways, and aside of different possible definitions of cluster to determine them for a large-scale network can be algorithmically challenging. Consequently, we find different solutions for one algorithm (if parameters in the algorithm are changed) and different solutions for different algorithms. One could call this an effect of the *choice of instrument for the measurement*. Last but not least, we can ask ourselves, if topics clearly delineated from each other really exist. Often in science very different topics still are related to each other. There exist unsharp boundaries and almost invisible long threads in the *fabric of science* (Boyack & Klavans, 2010), which might inhibit to find a contradiction-free solution. There is a seeming paradox between the fact that experts often can rather clearly identify what belongs to their field or a certain topic, and that it is so hard to quantitatively represent this with bibliometrics methods. However, a closer look into science history and science and technology studies reveals that what belongs to a field or a topic can still differ substantially also in the opinions of different experts; it changes over time; and even a defined canon or body of knowledge determining the essence of a field or a topic might be still subject to controversies and changes.

In the quest to define a topic two things collide. The principal, methodological and data-based ambiguity of what a topic is and the necessity to define a topic for purposes of education, knowledge acquisition and evaluation. This makes it such an intriguing problem to be solved. Because, different perspectives can be valid, there is also a need to preserve the above sketched diversity or ambiguity. Having said this, for the sake of scientific reasoning it is also necessary to be able to further specify the validity and appropriateness of different methods to define topics and fields. This paper contributes to the development of methods to compare algorithms and to visualize their different results.

We contribute to this sorting out process in two different ways. First, we apply standard clustering techniques to a specific article matrix built in a specific way from what we call a semantic matrix, in which rows are formed by entities from the bibliographic records of the articles (author names, journal ISSNs, topical terms, subjects, and other characteristics), columns by reduced dimensions from co-occurrence of entities and topical terms (one subset of the entities) over the whole set of articles. While we explain this in detail later, let us note here that the approach is conceptually more similar to classical information retrieval techniques based on Salton's vector space model than to usual bibliometrical mapping techniques (Salton, McGill 1983).

In a second step, we present an interactive visual interface called *LittleAriadne* that allows to display the context around those extracted and networked entities. The interface responds to a search query with a network visualization of most related terms, authors, journals and (other) cluster numbers. The query entry can be words, authors, but also cluster solutions. The displayed nodes or entities around a query term represent to a certain extend the context of this query. Depending on the query entry, we will see more or less other terms, journals, or authors. The interface allows to foreground one of entity types by selecting them. The interface has been originally developed for a much larger bibliographic database. In this paper our research questions are:

- Q1: How does the *Ariadne* algorithm work on a much smaller, field specific dataset? What possibility do we have to relate the produced contexts to domain knowledge?
- Q2: Can we use *Ariadne* to label the clusters produce by the different methods?
- Q3: Can we use *Ariadne* to compare different cluster assignments of papers, by treating those cluster assignments as additional entities? What can we visually learn about the topical nature of these clusters?

**Data**

The dataset used in this paper – called *Berlin dataset* - entails papers published in the period 2003-2010 in 59 astrophysical journals. Those papers have been downloaded from the Web of Science in the context of a German funded research project called "Measuring Diversity of Research," conducted at the Humboldt-University Berlin - hence the coined name *Berlin dataset*. It contains 120.007 records in total. Eventually, 111,616 records of the document types Article, Letter and Proceedings Paper have been treated with different clustering methods (see the other contributions for this special session).

Some of those cluster outcomes have been shared and are later displayed in the visual interactive interface. Table 1 shows the label of the different sets of clusters *x* we have included in *LittleAriadne*, whereby *x={a, b, ..., f}*. We have noted by which group cluster solutions were produced in the *Source* column. Each clustering method produced a set of clusters, whereby *y* stands for the number of clusters in a set. In our paper we used cluster solutions from CWTS (label: cwts 1.8), Cornell, Humboldt-University Berlin (hu), SciTech (sts-rg), KU Leuven (bc15) and one of our own (oclc_20). Except of cluster set *e*, they are all of the same order of magnitude. Because *Ariadne* relies on statistics across a corpus of articles as large as possible to produce semantic relatedness, we decided to discard clusters with less than 4 articles. But, from the solutions with many clusters (*d, e*) we decided to not display all. The last column in Table 1 gives the final numbers of the clusters from different clustering solutions.

**Table 1. Statistics of clusters generated from different methods.**

| x | Source | *y=#Cluster* | *#Cluster in Ariadne* |
|---|---|---|---|
| a | cwts 1.8 | 23 | 23 |
| b | cornell | 23 | 23 |
| c | oclc_20 | 20 | 20 |
| d | hu | 139 | 48 |
| e | sts-rg | 5664 | 229 |
| f | bc15 | 15 | 15 |

## Method

*Ariadne - an interactive visualization to navigate entities from large bibliographic databases*

The *Ariadne* algorithm has been developed on top of the article database, *ArticleFirst* of OCLC. The interface, accessible at http://thoth.pica.nl/relate, allows users to visually and interactively browse 35 thousand journals, 3 million authors, 1 million topical terms associated with 65 million articles (Koopman et al., 2015). For the purpose of this paper, we applied the same method on the Berlin dataset and built an instantiation, *LittleAriadne*, accessible at http://thoth.pica.nl/astro/relate.

**Table 2. An article from the Berlin dataset**

| | |
|---|---|
| Article ID | ISI:000276828000006 |
| Title | On the Mass Transfer Rate in SS Cyg |
| Abstract | The mass transfer rate in SS Cyg at quiescence, estimated from the observed luminosity of the hot spot, is log M-tr = 16.8 +/- 0.3. This is safely below the critical mass transfer rates of log M-crit = 18.1 (corresponding to log T-crit(0) = 3.88) or log M-crit = 17.2 (corresponding to the ""revised"" value of log T-crit(0) = 3.65). The mass transfer rate during outbursts is strongly enhanced |
| Author | [author:smak j] |
| ISSN | [issn:0001-5237] |
| Subject | [subject:accretion, accretion disks] [subject:cataclysmic variables] [subject:disc instability model] [subject:dwarf novae] [subject:novae, cataclysmic variables] [subject:outbursts] [subject:parameters] [subject:stars] [subject:stars dwarf novae] [subject:stars individual ss cyg] [subject:state] [subject:superoutbursts] |
| Cluster label | [cluster:a 19] [cluster:b 16] [cluster:c 15] [cluster:d 51] [cluster:e 17] [cluster:f 1] |

Table 2 shows for one example article from the *Berlin dataset* those fields of the bibliographic record that we used for *LittleAriadne*. It also shows which categories of entities we have. The ISI record ID has been used among the teams to compare solutions. For *Ariadne* as an interface it does not matter. *Ariadne* is different from a usual Information Retrieval search engine because it does not primarily deliver lists of documents matching a query, but a network of those entities which profile in the whole corpus ‘resonate’ most with the query entry. We come back to this aspect later. We further define so-called topical terms. Topical terms are frequent single or two-word phrases extracted from all titles and abstracts, for example, "mass transfer" and "quiescence" in our example. Next to the topical term, each author name is treated as an entity. In Table 2 we display the author name (and other entities below) in a syntax that can be used in the search field of the interface to search for a specific author. The next type of entities is the ISSN number of a journal. One can search for a single journal using the ISSN number, in the visual interface the journal title is used as label for a node representing a journal. Further, we have so-called subjects as separate entity type. Those subjects origin from the fields "Author Keywords" and "Keywords Plus" of the original Web of Science records. As last type of entities we add - and this is specific for *LittleAriadne* - to each of the articles cluster labels from their assignments to clusters produced by different teams. For example, the article in Table 2 has been assigned to cluster number 19 by source a (cwts 1.8) and number 16 by source b (cornell), and so on. In other words, we treat the cluster assignments of articles as they would be classification numbers or additional subject headings.

With the above detailed parsing of the bibliographic records we then build the matrix C (see Figure 1). In C frequent topical terms, and subjects, author names, cluster labels and journals appearing in the *Berlin dataset* form the rows and topical terms as well as subjects are listed in columns. The relatedness between all entities is computed based on the *context* they share, instead of direct co-occurrences in the data. The context of these entities is captured by their co-occurrences with topical terms and subjects, that is, we count how often an author, or a cluster label co-occurs with a certain topical term or subject in an article, summing up over all articles in the corpus. In the Berlin dataset, we have in total 90,343 entities, including 59 journals, 27,027 author names (single instances, no author disambiguation applied), 358 cluster IDs, 39,577 topical terms and 23,322 subjects. This would produce a sparse matrix of roughly 90K x 63K that is expensive for computation.

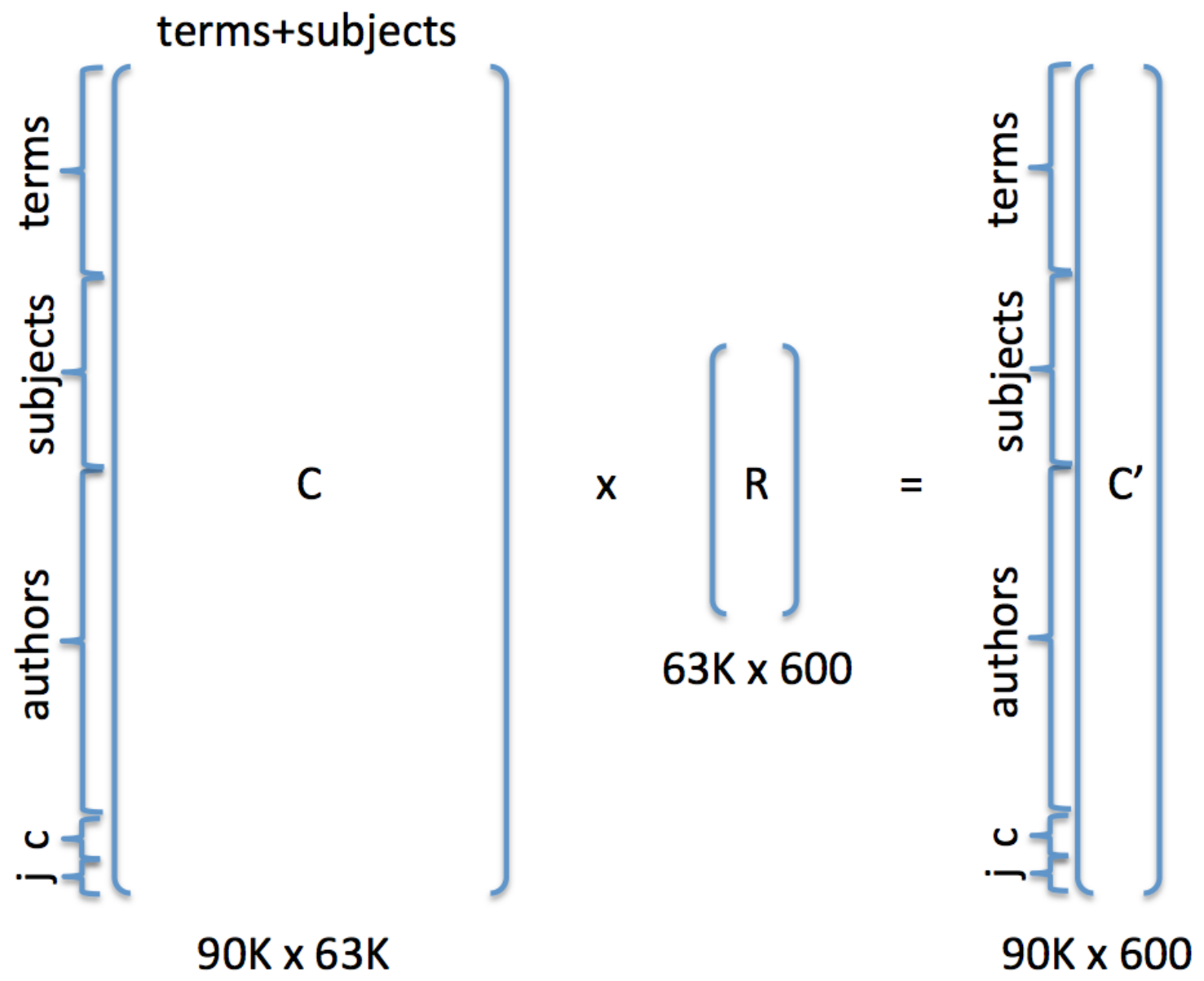

**Figure 1. Dimension reduction using Random Projection**

To make the algorithm scale and produce a responsive visual interface, we applied *Random Projection* (Johnson & Lindenstrauss 1984, Achlioptas 2003) to reduce the dimensionality of the matrix. As shown in Figure 1, by multiplying C with a 63K x 600 matrix of randomly distributed -1 and 1, the original 90K x 63K matrix C is reduced to a *Semantic Matrix C'* of the size of 90K x 600, with each row vector representing the semantics of an entity. With this Semantic Matrix, the interactive visual interface dynamically computes the most related entities (e.g. ranked by cosine similarity) to a search query and presents a networked visualization of the context of a query term whereby entities are positioned closer to each other if they are more related to each other.

*OCLC clusters production - Clustering the Berlin dataset using the Semantic Matrix*
The Ariadne interface provides a networked view about entities associated with articles, but it does not produce article clusters straightaway. In order to cluster articles, we need to build a semantic representation of each article. We receive the semantic representation for an article by the following steps. For each article, we lookup all entities related to this article in the Semantic Matrix C'. For our example in Table 2 we have one vector representing the single author of that article in the whole Semantic Matrix, 12 vectors representing the subjects, one vector for the journal, 6 vectors representing the cluster labels and *n* vectors for all extracted topical words. In other words, each article is represented by a subset of vectors and the vector components correspond to the dimensions of the Semantic Matrix. We then take the average of those single entity vectors as the semantic representation of a specific article. All articles together form a matrix M with 111,616 rows and 600 columns. We applied a standard clustering technique - the MiniBatchKmeans method (Sculley 2010) - to M. We used the scikit-learn python library (http://scikit-learn.org/) for this. Applied to the *Berlin dataset* we receive a cluster solution with a comparable size of k=20 clusters, labeled as oclc_20, and a unique assignment of articles to this cluster.

**Results - The Berlin set in *LittleAriadne***
We used the visual, interactive interface built for the *Berlin dataset* to the context around a specific cluster solution and the similarity between different ones. For this we performed different experiments, which correspond to the research questions Q1-Q3 of the introduction

- Experiment 1: We used *LittleAriadne* as information retrieval tool. We searched with query terms, inspected and navigated through the resulting network visualization. (Q1)
- Experiment 2: We used the semantic matrix to provide the most related topical terms for each cluster as an approximation of cluster labels. (Q2)
- Experiment 3: We used the query syntax to display two or more cluster solutions together in one overview. (Q3)

*Experiment 1 - Information retrieval*

In *LittleAriadne* we can now study the *Berlin dataset* as any other dataset. Figure 2 gives a snapshot of the context about "magnetic flux" used as query term. The most related topical terms and subjects are shown, together with 3 most related clusters provided by CWTS, Cornell and SciTech (coded in different colors). Each node is clickable which leads to another visualization of the context of the selected node. When mousing over a node, one sees how often this entity occurs in the whole corpus. Given that different statistical methods are at the core of the Ariadne algorithm, this give an indication of the reliability of the suggested position and links. In the interface one can further refine the display. For instance, one can choose the number of nodes to be shown or decide to limit the display to only authors, journals, topical terms or clusters. Within the interface, one can navigate the context of entities in the *Berlin dataset* by seamlessly travelling between authors, journals, topical terms and clusters in a visual and interactive way.

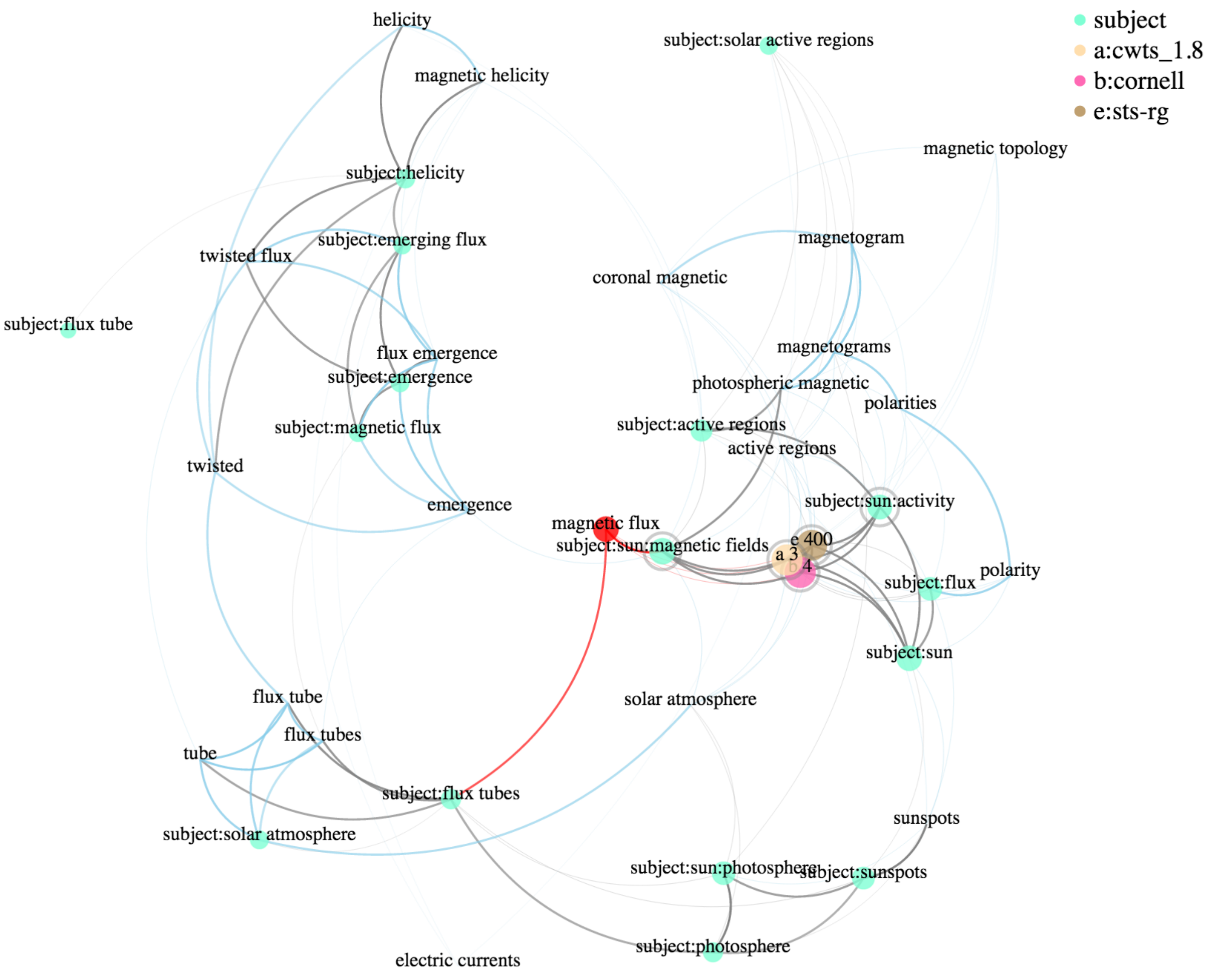

**Figure 2. Context around "magnetic flux"**[1]

*Experiment 2 -Labeling clusters*

Please note, that in *LittleAriadne* we cannot see the position of articles in relations to the different entities. One could say that the articles produce the elements of the networked context, but they themselves are distributed over it. What we can do is to switch to a view that shows most related topical terms, subjects, journals, authors, and other clusters. The outcome of such a *click-through* action is shown in Figure 3. In this example, the most related topical terms, subjects, one journal, and four other clusters are presented as the contextual information about the cluster "a 2".

It is now possible to label each clusters using the most related topical terms. As shown in Table 3, the 9 topical terms most related to cluster "a 2" are "cosmology," "dark energy," "density perturbations," "cosmologies," "planck," "cosmological," "spatial curvature," "inflationary," and "inflation." Together they give a rough idea about what this cluster with 8954 articles is about, but it requires domain expertise to evaluate and transform them into real cluster labels, meaning representing names of specialties, topics or fields used by the scientific community, a well-known problem of bibliometric mapping (Noyons, 2005).

[1] Accessible at http://thoth.pica.nl/astro/relate?input=magnetic+flux.

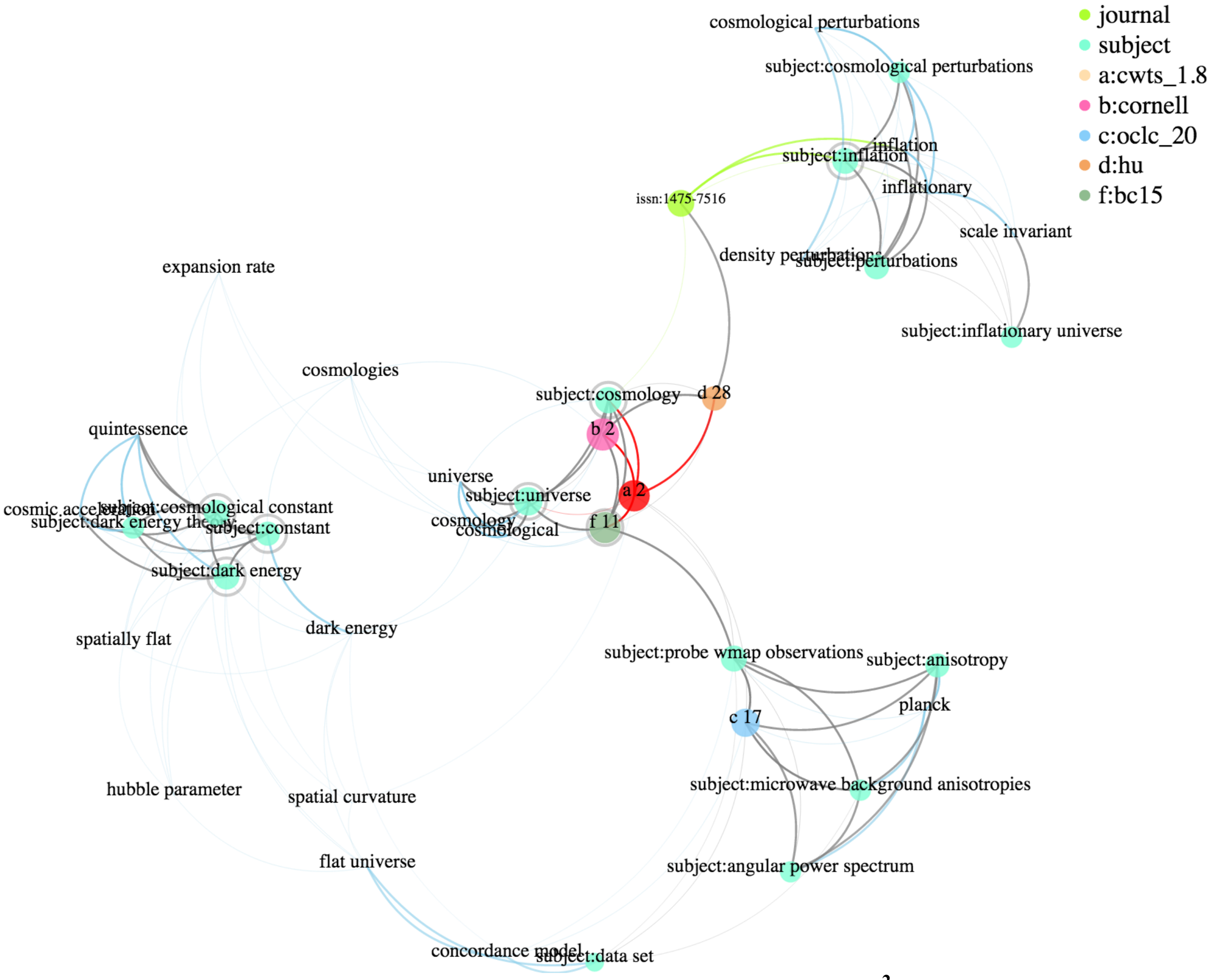

**Figure 3. The contextual view of cluster "a 2"[2]**

**Table 3. Top related topical terms**

| Cluster ID | Top 9 most related topical terms |
|---|---|
| a 2 | "cosmology" "dark energy" "density perturbations" "cosmologies" "planck" "cosmological" "spatial curvature" "inflationary" "inflation" |
| b 2 | "cosmology" "cosmological constant" "cosmologies" "cosmological" "universes" "dark energy" "quadratic" "tensor" "planck" |
| c 17 | "power spectrum" "cosmological parameters" "cmb" "last scattering" "anisotropies" "microwave background" "power spectra" "planck" "cosmic microwave" |
| d 28 | "density perturbations" "inflationary" "inflation" "dark energy" "scale invariant" "spatial curvature" "cosmological perturbations" "inflationary models" "cosmologies" |
| f 11 | "cosmology" "cosmological" "dark energy" "universe" "planck" "density perturbations" "cosmologies" "spatial curvature" "flat universe" |

[2] Accessible at http://thoth.pica.nl/astro/relate?input=%5Bcluster%3Aa+2%5D.

*Experiment 3 - Comparing cluster solutions*

In *LittleAriadne* we extended the interface with a possibility to compare sets of clusters. In Figure 4 (a) we can visually see the high similarity between clusters from CWTS and those from Cornell. Nearly each CWTS cluster is accompanied by a Cornell cluster. Figure 4 (b) shows two other sets of clusters which partially agree with each other but also clearly have different capacity in distinguishing different clusters. Figure 5 shows all the cluster entities from all six clustering solutions. Given the amount of the clusters, it is difficult to grasp the detailed difference between solutions. However, this visualization does provide a general overview of all the clustering solutions, based on their similarities to each other.

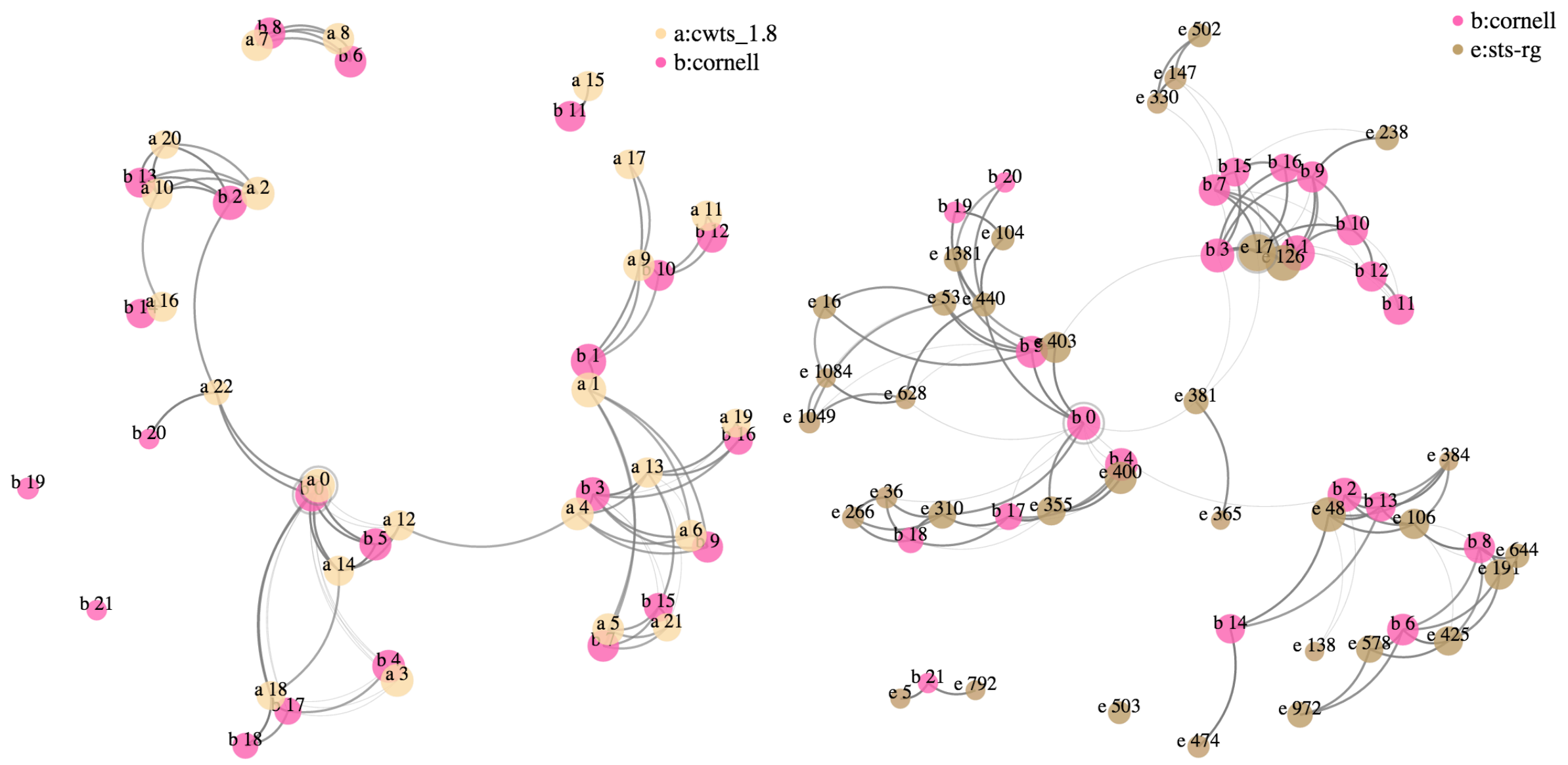

**(a) Highly similar (between CWTS 1.8 and Cornell)[3]**

**(b) Partially agreeing (between Cornell and SciTech)[4]**

**Figure 4. Comparison between sets of clusters**

## Discussion and conclusion

We present a method and an interface that allows browsing through the contexts of entities, such as topical terms, authors, journals and subjects associated with a set of documents. We have applied the method to the problem of topic delineation addressed in this special session. Because the tool shows (local) context and not the position of single documents in relation to clusters we think it has a potential to be complementary to any other method of cluster comparison. In particular we have asked how the *Ariadne* algorithm works on a much smaller, field specific dataset? Not surprisingly, compared with our exploration in the ArticleFirst interface, we find more consistent representations. That means that specific vocabulary is displayed, which can be cross-checked in Wikipedia or Scholar Google, for which the interface offers a direct click through.

---

[3] Accessible at http://thoth.pica.nl/astro/relate?input=%5Bcluster%3Aa%5D%5Bcluster%3Ab%5D&type=S&show=50.

[4] Accessible at http://thoth.pica.nl/astro/relate?input=%5Bcluster%3Ae%5D%5Bcluster%3Ab%5D&type=S&show=300.

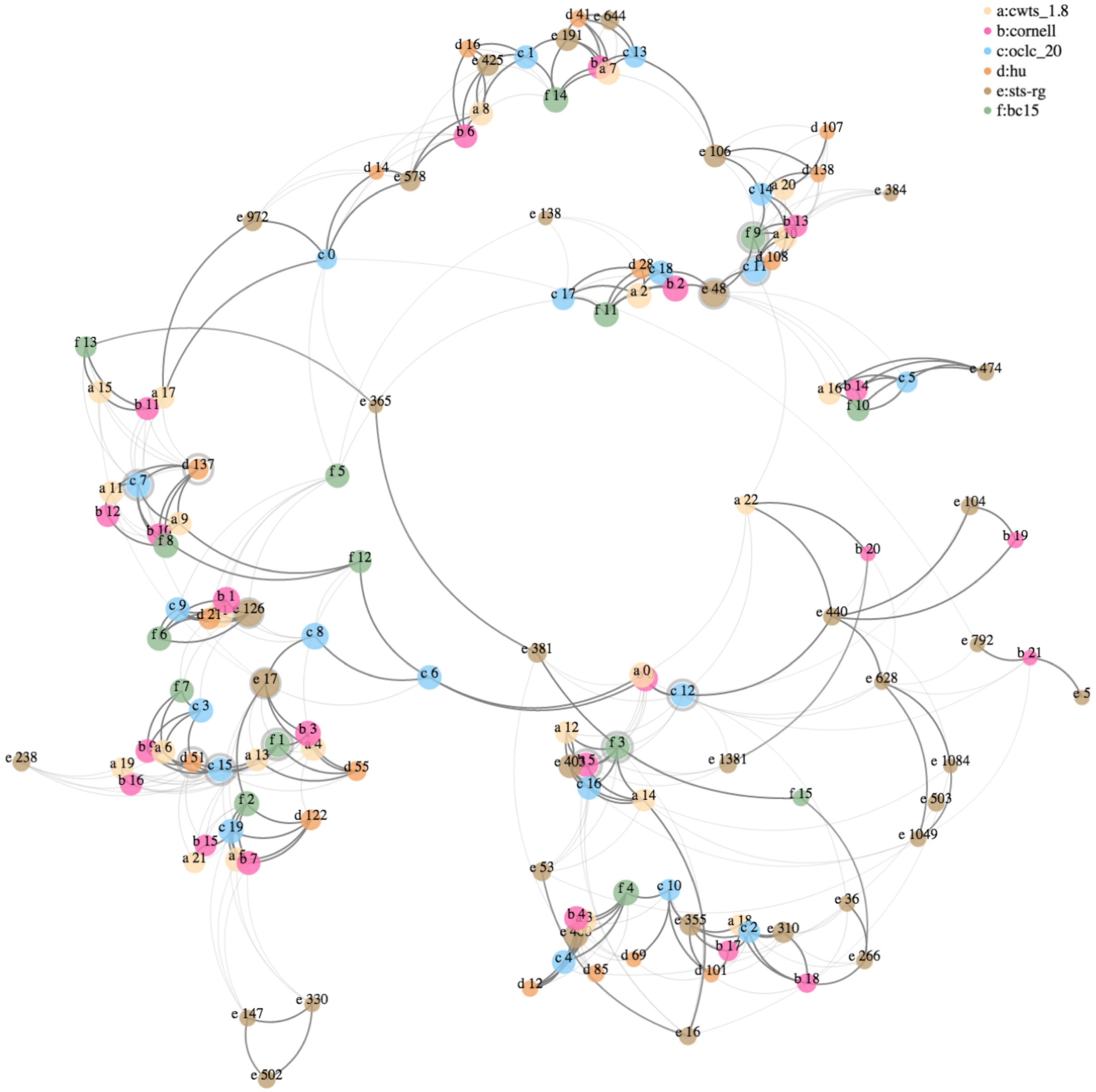

**Figure 5. Comparing clusters from 6 clustering solutions**

On the other hand, the bigger number of topical terms in the larger database leads to a situation where almost every query term produces a response. In *LittleAriadne* searches for e.g. literary persons as Jane Austen retrieve nothing - a blank screen. In preparation of this paper we surfed through the interface, and compared the most relevant topical terms around a cluster to other classifications used in Astrophysics, such as Physics and Astronomy Classification Scheme (PACS®[5]). In this punctual exploration we did find correlations between the names of PACS classes (subclasses, and related controlled vocabulary) and the selected topical terms in *LittleAriadne*. We will further compare the context around clusters and the suggested related topical terms with labels produced by other teams in this special session. Ultimately, the discussion with domain experts belongs to a proper evaluation of the interface. We demonstrated that we can use *LittleAriadne* to compare different cluster solutions mutually and even generate a wider overview. We will discuss in the special session how Ariadne can further be of use in the comparison of clustering and delineation of topics. At least, we hope that this interactive tool supports discussion about different clustering algorithms and helps to find the right meaning of clusters, and appropriate labels for them.

[5] http://www.aip.org/publishing/pacs/pacs-2010-regular-edition

We also have plans to further develop the Ariadne algorithm. The Ariadne algorithm is general enough to accommodate additional types of entities to the semantic matrix. In the future, we plan to add citations, publishers, conferences, etc. with the aim to provide a richer contextualization of entities. We also plan to add links to articles that contribute to the contextual visualization, this way strengthening the usefulness of *Ariadne* not only for the associative exploration of contexts similar to scrolling through a systematic catalogue, but also as a direct tool for document retrieval. In this context we plan to further compare *LittleAriadne* and *Ariadne*. In a first attempt, we 'projected' the astrophysical documents into *ArticleFirst* by looking them up in the large semantic matrix built for Ariadne. We found the resulting representations less consistent when browsing through. That is not a surprise, because when merging them you see how field-specific content fits and miss-fits into many other contextualizations. The advantage of *LittleAriadne* is the confinement of the dataset to one scientific field and topics within. We hope by continuing such experiments also to learn more about the relationship between genericity and specificity of contexts, and how that can be best addressed in information retrieval.

**Acknowledgments**

Part of this work has been funded by the COST Action TD1210 KnoweScape.